\documentclass[twocolumn,aps,prl,showpacs,superscriptaddress]{revtex4}
\usepackage{epsfig,graphics}
\usepackage{graphicx}
\usepackage{dcolumn}
\usepackage{bm}
\usepackage{amsmath}
\usepackage[usenames]{color}
\usepackage{ulem} 

\begin{document}

\title{Time evolution of Mach-like structure in a partonic transport model}

\author{ G. L. Ma}
\affiliation{Shanghai Institute of Applied Physics, Chinese
Academy of Sciences,   Shanghai 201800, China}
\author{ S. Zhang}
\affiliation{Shanghai Institute of Applied Physics, Chinese
Academy of Sciences,   Shanghai 201800, China}
\affiliation{Graduate School of the Chinese Academy of Sciences,
Beijing 100080, China}
\author{ Y. G. Ma}
\thanks{Corresponding author: Email: ygma@sinap.ac.cn}
\affiliation{Shanghai Institute of Applied Physics, Chinese
Academy of Sciences,   Shanghai 201800, China}
\author{ X. Z. Cai}
\affiliation{Shanghai Institute of Applied Physics, Chinese
Academy of Sciences,   Shanghai 201800, China}
\author{ J. H. Chen}
\affiliation{Shanghai Institute of Applied Physics, Chinese
Academy of Sciences,   Shanghai 201800, China}
\affiliation{Graduate School of the Chinese Academy of Sciences,
Beijing 100080, China}
\author{ Z. J. He}
\affiliation{Shanghai Institute of Applied Physics, Chinese
Academy of Sciences,   Shanghai 201800, China}
\author{H. Z. Huang}
\affiliation{ University of California at Los Angeles, CA 90095,
USA}
\author{ J. L. Long}
\affiliation{Shanghai Institute of Applied Physics, Chinese
Academy of Sciences,   Shanghai 201800, China}
\author{ W. Q. Shen}
\affiliation{Shanghai Institute of Applied Physics, Chinese
Academy of Sciences,   Shanghai 201800, China}
\author{ X. H. Shi}
\affiliation{Shanghai Institute of Applied Physics, Chinese
Academy of Sciences,   Shanghai 201800, China}
\affiliation{Graduate School of the Chinese Academy of Sciences,
Beijing 100080, China}
\author{ C. Zhong}
\affiliation{Shanghai Institute of Applied Physics, Chinese
Academy of Sciences,   Shanghai 201800, China}
\author{ J. X. Zuo}
\affiliation{Shanghai Institute of Applied Physics, Chinese
Academy of Sciences,   Shanghai 201800, China}
\affiliation{Graduate School of the Chinese Academy of Sciences,
Beijing 100080, China}

\date{ \today}

\begin{abstract}
The time evolution of Mach-like structure (the splitting of the
away side peak in di-hadron $\Delta\phi$ correlation) is presented
in the framework  of a dynamical partonic transport model. With
the increasing of the lifetime of partonic matter, Mach-like
structure can be produced and developed by strong parton cascade
process. Not only the splitting parameter but also the number of
associated hadrons ($N_{h}^{assoc}$) increases with the lifetime
of partonic matter and partonic interaction cross section. Both
the explosion of $N_{h}^{assoc}$ following the formation of
Mach-like structure and the corresponding results of
three-particle correlation support that a partonic Mach-like shock
wave can be formed by strong parton cascade mechanism. Therefore,
the studies about Mach-like structure may give us some critical
information, such as the lifetime of partonic matter and
hadronization time.

\end{abstract}


\maketitle

Recent RHIC experimental results indicated an exotic partonic
matter may be created in central Au + Au collisions at
$\sqrt{s_{NN}}$=200 GeV. When a parton with high transverse
momentum (jet) passes through the new matter, it was predicted
that jet will quench and lose energy, i.e. jet
quenching~\cite{HIJING}. At the same time, the lost energy will be
redistributed into the medium \cite{soft-soft-th1,Vitev}.
Experimentally the soft scattered particles which carry the lost
energy have been reconstructed via di-hadron angular correlations
of charged particles \cite{soft-soft-ex}. It is very interesting
that Mach-like structure (the splitting of the away side peak in
di-hadron $\Delta\phi$ correlation) has been observed in di-hadron
azimuthal correlations in central Au + Au collisions at
$\sqrt{s_{NN}}$ = 200 GeV \cite{soft-soft-ex,sideward-peak2}. It
was proposed that the structure is due to a Mach-cone shock wave
generation, because jets travel faster than the sound in the new
medium \cite{Stocker,Casalderrey}.  However a gluon Cherenkov-like
radiation model can also produce such  structure
\cite{Vitev,Igor,Koch}. Though so far some
publications~\cite{Ruppert,Thorsten,Chaudhuri,Sat} based on above
different ideas come forth, quantitative understanding of the
experimental observation has yet to be established.

In this work, a dynamical multi-phase transport model
(AMPT)~\cite{AMPT}, that includes both initial partonic and final
hadronic interactions, will be used to study the production
mechanism of  Mach-like structure. AMPT model is a hybrid Monte
Carlo model which consists of four main processes: the initial
conditions, partonic interactions, the conversion from partonic
matter into hadronic matter and hadronic rescattering. The initial
conditions, which include the spatial and momentum distributions
of minijet partons and soft string excitations, are obtained from
the HIJING model \cite{HIJING}. Excitations of strings melt
strings into partons. Scatterings among partons are modelled by
Zhang's parton cascade model   \cite{ZPC}, which at present
includes only two-body scattering with cross section obtained from
the pQCD with screening mass. In  the string melting version of
the AMPT model (we briefly call it as "the melting AMPT"
model)~\cite{SAMPT}, a simple quark coalescence model is used to
combine partons into hadrons when all partons stop interactions.
Dynamics of the subsequent hadronic matter is then described by a
relativistic transport model. 
Details of the AMPT model can be found in a recent review
\cite{AMPT}. It has been shown that in previous studies the
partonic effect can not be neglected and the string melting
mechanism is appropriate when the energy density is much higher
than the critical density for the QCD phase transition
\cite{AMPT,SAMPT,Jinhui}. In previous AMPT studies, parton cascade
process will stop until partons do not interact again, and the
hadronization takes place dynamically during the process. However,
it is not very reasonable since the lifetime of partonic matter
should be limited and decided by when the energy density
(temperature) of reaction system enters into the critical value,
in the view of Lattice QCD calculations\cite{LatticeQCD}. In the
present work, we will test different lifetimes of partonic matter,
after which no partonic interactions are allowed and all left
partons must coalesce into hadrons suddenly. It means that the
partonic matter come to the critical point at a certain time,
therefore the lifetime of partonic matter could be relative to
hadronization time of reaction system. Two partonic cross
sections, 10mb and 3mb, were used in our work.

Di-hadron correlations between the trigger hadrons and the
associated ones were constructed by a  mixing-event technique in
our analysis, as experimenters did. The  $p_{T}$ window cuts for
trigger and associated particles are $2.5 < p_{T}^{trig} < 4$
GeV/$c$ and $1.0 < p_{T}^{assoc} < 2.5$ GeV/$c$. Both trigger and
associated particles are selected within pseudo-rapidity window
$|\eta| < 1.0$. The pairs of the associated particles with trigger
particles in same events are accumulated to obtain $\Delta\phi =
\phi - \phi_{trig}$ distributions. In order to remove the
background which is expected to mainly come from the effect of
elliptic flow \cite{soft-soft-ex,sideward-peak2}, so-called
mixing-event method is applied to simulate the background. In this
method, we mixed two events which have very close centrality into
a new mixing event, and extracted $\Delta\phi$ distribution which
is regarded as the corresponding background. A zero yield at
minimum assumption is adopted to subtract the background as did in
experimental analysis \cite{sideward-peak2,di-hadron}.

\begin{figure}[htbp]
\resizebox{3in}{3in}{\includegraphics[0in,0in][6in,7in]{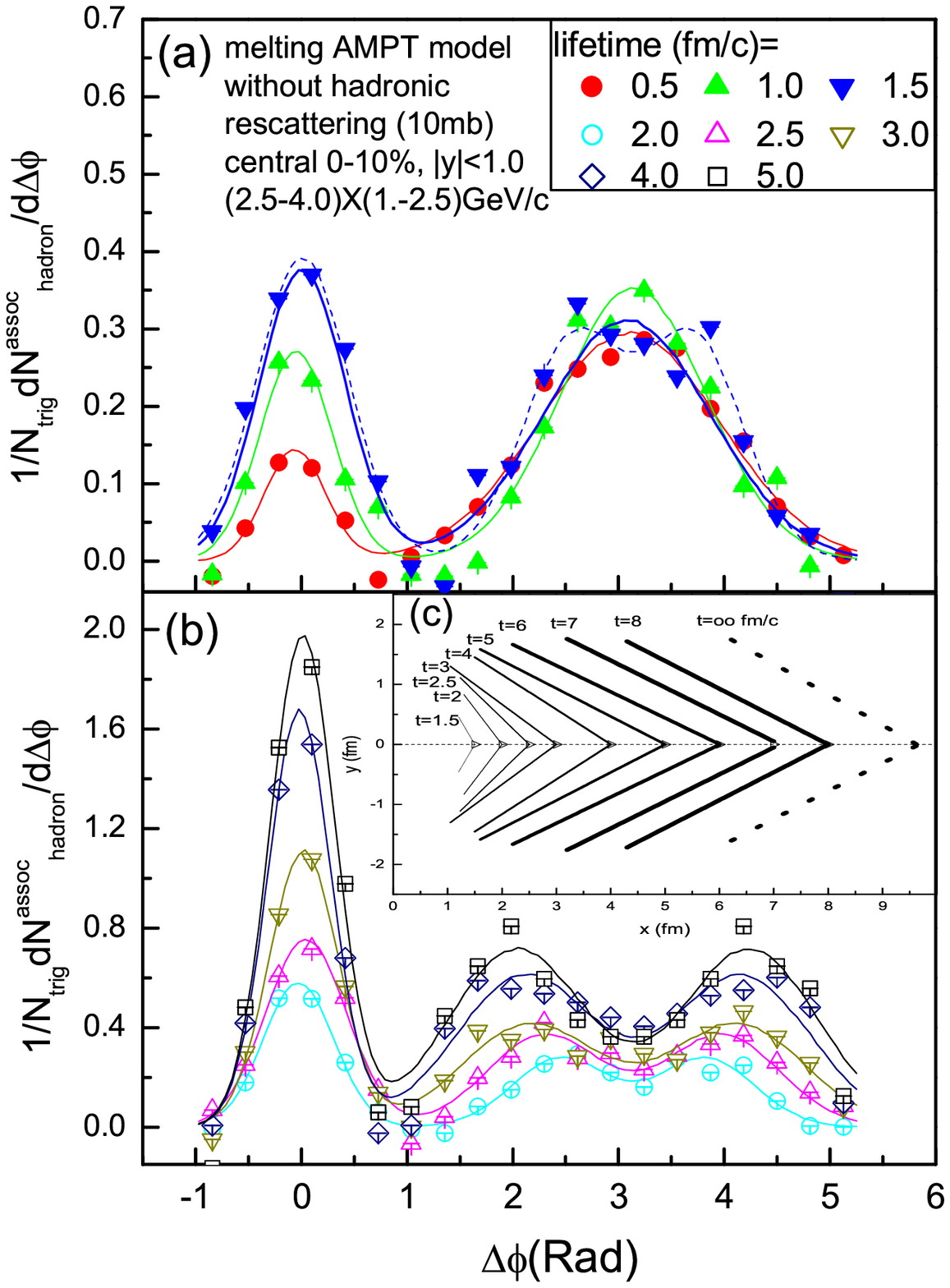}}
\vspace*{-0.2cm} \caption{(Color online) The $\Delta\phi$
correlations between trigger hadrons and associated ones ($2.5 <
p_{T}^{trig} < 4.0 $GeV/$c$ and $1.0 < p^{assoc}_T < 2.5 $
GeV/$c$) in Au + Au collisions at $\sqrt{s_{NN}}$ = 200 GeV for
different lifetimes of partonic matter in the melting AMPT model
(10 mb) without hadronic rescattering. (a): lifetime = 0.5, 1.0,
1.5 fm/c; (b): lifetime = 2.0, 2.5, 3.0, 4.0, 5.0 fm/c. The
$\Delta\phi$ correlations at different lifetimes have been shown
with different makers for clarity, and corresponding solid lines
are their two-Gaussian (a) or three-Gaussian (b) fits. (The dash
line in (a) shows a three-Gaussian fit to the case of lifetime =
1.5 fm/c.); The inserted (c) in (b) shows a schematic illustration
of Mach-like structure where the patonic Mach-like shock wave was
excited by a jet travelling from x=0 fm in the positive x
direction. (The different lines represent the partonic Mach-like
shock fronts at different times.)}\label{fig1}
\end{figure}

Figure~\ref{fig1} shows di-hadron $\Delta\phi$ correlations in Au
+ Au collisions at $\sqrt{s_{NN}}$=200 GeV for different lifetimes
of partonic matter in the melting AMPT model without hadronic
rescattering. In figure~\ref{fig1}(a), no Mach-like structure is
observed within $\sim$ 1.0 fm/c partonic lifetime. However,
Mach-like structure appears by longer parton cascade processes
(more than $\sim$ 1.5 fm/c), as shown in Fig.~\ref{fig1}(a) and
(b). (Note: we give two types of fits to 1.5 fm/c for the
comparation.). It indicates that the formation of Mach-like
structure needs a long duration of interacting partonic phase.
Such partonic lifetime dependence is consistent with our previous
results that the splitting parameter $D$ (half distance between
two Gaussian peaks on away side of di-hadron $\Delta\phi$
distributions) increases with number of participants, which is due
to more long-lived partonic matters in the central collisions than
in the peripheral collisions \cite{di-hadron}.

\begin{figure}[htbp]
\resizebox{2.8in}{2.6in}{\includegraphics[0in,0in][7in,7in]{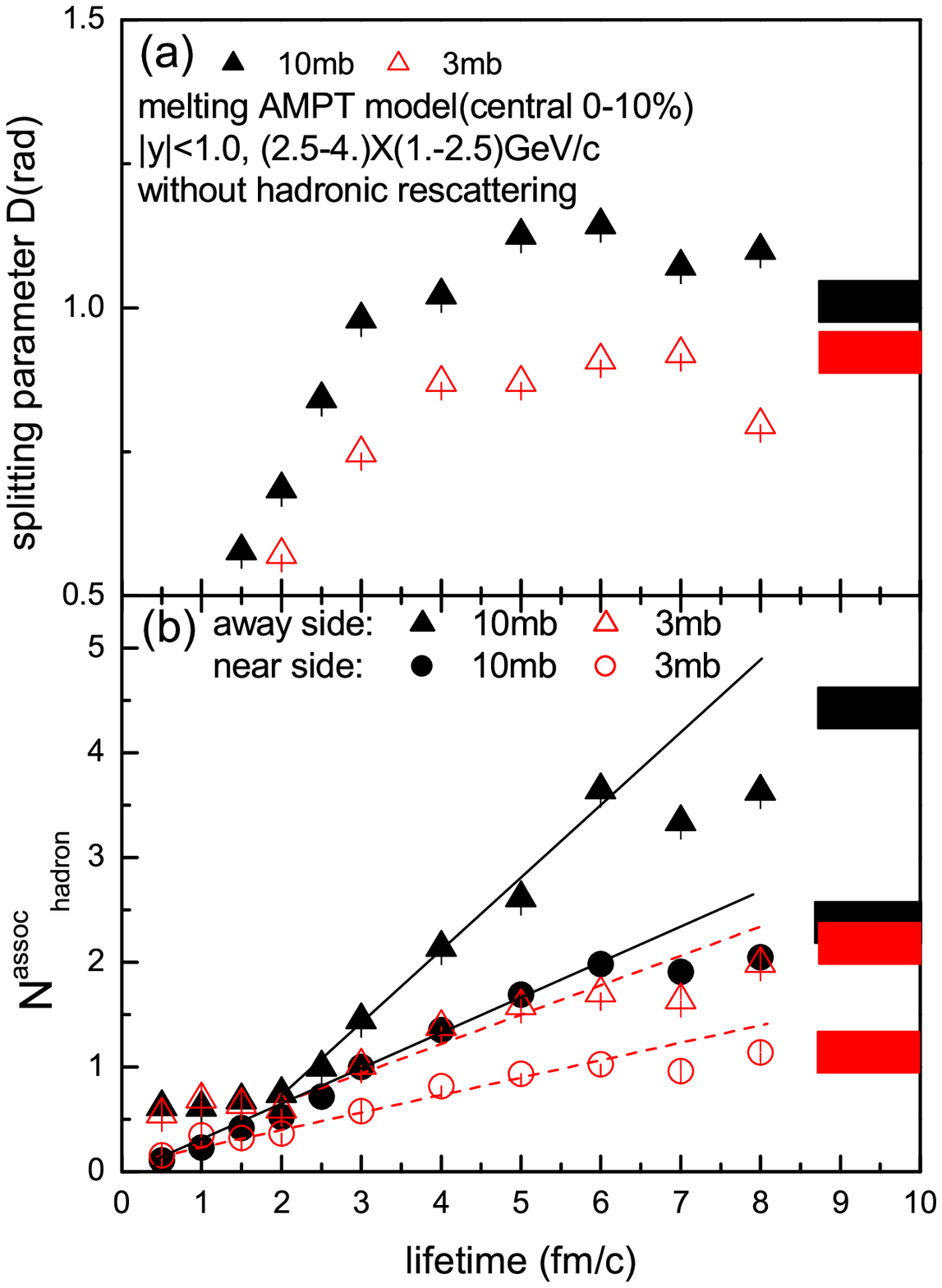}}
\vspace*{-0.5cm}
    \caption{(Color online) The dependences of splitting parameter D (a) and number of associated
    hadrons (b) on the lifetime of partonic matter in Au + Au collisions at $\sqrt{s_{NN}}$=200 GeV in the melting
    AMPT model without hadronic rescattering. Full points are for 10mb and open ones for 3mb. Triangles are
    for away side and circles for near side. Lines give the linear fit functions for $N_{h}^{assoc}$ vs lifetime (solid: 10mb and dash: 3mb).
    Bands show the corresponding values at an infinite partonic lifetime, i.e. the results in our previous AMPT set~\cite{di-hadron}.}
    \label{fig2}
\end{figure}

Quantitatively, Figure~\ref{fig2} gives the dependences of
splitting parameter $D$ on the lifetime of partonic matter. It is
observed that the splitting parameter $D$ increases and then
saturates with the lifetime of partonic matter, and the result
with 10mb gives  much bigger splitting parameters than that with
3mb, in comparison with experimental results. (But the opening
angle of Mach-like structure can only be observed after the
lifetime of 2 fm/c for 3mb.) It indicates that the production and
development of Mach-like structure needs a long-lived and strong
parton cascade process. It should be pointed out that the time
evolution of energy density and d-quark temperature have been
shown to reach the critical value around $\sim$ 5 fm/c in previous
AMPT works~\cite{lwchenphiOmega}. The parton elliptic flow also
saturates at this time in parton cascade evolution~\cite{SAMPT}.
Here the corresponding saturation time of splitting parameter D is
consistent with them. The hydrodynamical works
\cite{Casalderrey,Thorsten}, which apply Cooper-Frye method for
hadronization at some freeze-out temperature, have given good
descriptions about Mach-like structure. Therefore, though partonic
hadronization may be not an instantaneous phase transition but a
durative process, the researches on Mach-like structure may shed
light on some information about the lifetime of partonic matter
and hadronization time. On the other hand, both numbers of
associated hadrons ($N_{h}^{assoc}$)  on near side and away side
increase with lifetime (Figure~\ref{fig2}(b)). It is remarkable
that the production rate of associated hadrons
($dN_{h}^{assoc}$/dt, the slope of $N_{h}^{assoc}$ vs lifetime)
has two different values before and after birth of Mach-like
structure on away side, while that of associated hadrons on near
side always keeps a constant. The Table \ref{tab:slope} gives the
production rates of associated hadrons on near side and away side
in Au + Au collisions at $\sqrt{s_{NN}}$=200 GeV in the melting
AMPT model without hadronic rescattering, which indicates a
Mach-like shock wave has been created by strong parton cascade
process and then its birth will boost its own growth quickly,
therefore $'shock$ $partons'$ may be ejected from the shock wave
plentifully, especially for the case of bigger partonic cross
sections. The inserted (c) in figure~\ref{fig1} give a schematic
illustration of formation and development of the partonic
Mach-like shock wave.

\begin{table}[h]
\caption{\label{tab:slope}Production rates of associated hadrons
on near side and away side in Au + Au collisions at
$\sqrt{s_{NN}}$=200 GeV in the melting AMPT model without hadronic
rescattering.}
\begin{ruledtabular}
\begin{tabular}{cccc}
  $dN_{h}^{assoc}$/dt & near side  & away side & away side\\
                     & ($\leq$ 6 fm/c) & ($\leq$ 2 fm/c) & (2 $\sim$ 6 fm/c)\\ \hline
  10 mb& 0.35 $\pm$ 0.01 & 0.12 $\pm$ 0.01 & 0.69 $\pm$ 0.01 \\
  3 mb & 0.18 $\pm$ 0.01 & 0.10 $\pm$ 0.01 & 0.26 $\pm$ 0.01 \\
\end{tabular}
\end{ruledtabular}
\end{table}

\begin{figure}[htbp]
\resizebox{3.3in}{4.2in}{\includegraphics[0in,0in][8in,10in]{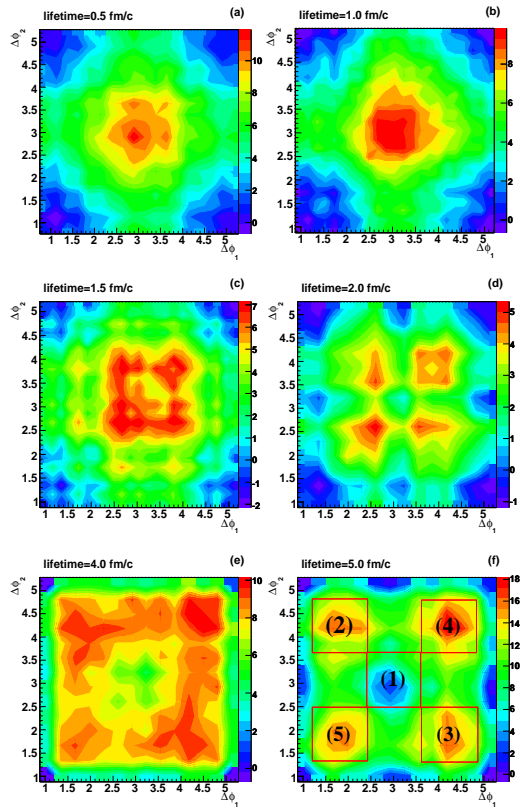}}
\vspace*{-0.5cm}
    \caption{(Color online) Background subtracted segmental 3-particle
     correlation areas ($1<\Delta\phi_{1,2}<5.28$) in central Au+Au collisions (0-10\%) at $\sqrt{s_{NN}}$ = 200 GeV
     with different lifetimes of partonic matter in the melting AMPT model(10 mb) without hadronic rescattering.
     }
    \label{fig3}
\end{figure}

The mixing-event technique has been used in our three-particle
correlation analysis~\cite{three-particle}. The $p_{T}$ window
cuts for trigger and associated particles are selected as  $2.5 <
p_{T}^{trig} < 4$ GeV/$c$ and $1.0 < p_{T}^{assoc} < 2.5$ GeV/$c$,
respectively. Both trigger and associated particles are selected
within pseudo-rapidity window $|\eta| < 1.0$. In the same events,
raw 3-particle correlation signals in $\Delta\phi_{1} = \phi_{1} -
\phi_{trig}$ versus $\Delta\phi_{2} = \phi_{2} - \phi_{trig}$ are
accumulated. Three background contributions are expected to be in
the raw signal. The first one is $'hard-soft'$ background which
comes from a trigger-associated pair combined with a background
associated particle. It can produced by mixing a
trigger-associated pair with another fake associated particle that
is from different event. The second one is $'soft-soft'$
background which comes from an associated particle pair combined
with a background trigger particle. It can be mimic by mixing an
associated particle pair with another fake trigger particle that
is from different event. The third one is a $'random'$ background,
which are produced by mixing a trigger particle and two associated
particles respectively from three different events. We require
that the mixed events are from very close centralities which can
be determined by impact parameter. In order to subtract the
background from the raw signals, we normalize the strip of
$0.8<|\Delta\phi_{1,2}|<1.2$ to zero.

Figure~\ref{fig3} gives the background subtracted segmental
3-particle correlation areas ($1<\Delta\phi_{1,2}<5.28$) in
central Au+Au collisions (0-10\%) at $\sqrt{s_{NN}}$ = 200 GeV,
which shows three-particle correlations among one trigger particle
and two associated particles on away side. There are three kinds
of three-particle correlations in the area. Let us take segmental
3-particle correlation area with lifetime = 5.0 fm/c
(Figure~\ref{fig3}(f)) as an example. The first one is $'center'$
region ($|\Delta\phi_{1,2}-\pi|<0.5$, i.e. region (1)) where
three-particle correlation mainly comes from one trigger particle
and two associated particles in the center of away side. The
$'center'$ correlations show penetration ability of away jet. The
second one is $'cone'$ region ($|\Delta\phi_{1}-(\pi\pm1)|<0.5$
and $|\Delta\phi_{2}-(\pi\mp1)|<0.5$, i.e. region (2) and region
(3)) where three-particle correlation gives the correlation
between one trigger particle and two associated particle which are
respectively from different cones of away side. It was predicted
that $'cone'$ correlations can  be caused by Mach-cone shock wave
effect when a jet goes faster than sound in the medium, shock wave
would appear on away side. The third one is $'deflected'$ region
($|\Delta\phi_{1,2}-(\pi\pm1)|<0.5$, i.e. region (4) and region
(5)) where three-particle correlation can reflect the correlation
between one trigger particle and two associated particles from the
same one of two peaks on away side in di-hadron $\Delta\phi$
correlations, which may be due to the sum of away-side jets
deflected by radial flow and Mach-cone shock wave effect. It is
found that there is only $'center'$ correlation before 1.5 fm/c,
but $'cone'$ and $'deflected'$ correlations appear afterward and
extend with the increasing of lifetime. Because the correlations
within one cone-peak and between two cone-peaks on away side in
di-hadron $\Delta\phi$ correlations are equivalent in a Mach-like
shock wave scenario, the corresponding three-particle correlation
strengths at ($\pi$-D,$\pi$-D), ($\pi$+D,$\pi$+D),
($\pi$-D,$\pi$+D) and ($\pi$+D,$\pi$-D) should be same. Actually,
both $'cone'$ and $'deflected'$ correlations have similar strength
in our results, which is consistent with the Mach-cone shock wave
mechanism.

As we know, the AMPT model with the string melting scenario has
presented good results of hadronic elliptic flow and even given
the mass ordering of elliptic flow~\cite{SAMPT,Jinhui} which has
been well described by hydrodynamics model. It can be attributed
to the big cross section of parton interaction in the AMPT model
which leads to strong parton cascade that couple the partons
together, it therefore induce the onset of hydrodynamics
behavior~\cite{Zhang99}. Our present results show that the
abundant and sequential partonic interactions can couple many
partons together to exhibit the other collective behavior, i.e.
partonic Mach-like shock wave \cite{di-hadron,three-particle}. In
Refs~\cite{Ruppert,Thorsten}, it was claimed that the main origin
of Mach-cone structures is the lost energy which is deposited into
a collective mode, i.e. $'hydro-mode'$. At the same time, we
noticed that no Mach-like cone in di-hadron $\Delta\phi$
correlations has been seen in a dynamical hydrodynamics model in
Ref.~\cite{Chaudhuri}. It should be mentioned that there are big
differences between two dynamical models. The linearized
hydrodynamical approximation is not applicable near jet region
where the medium is with rapid variation of energy density and
without adequate thermalization~\cite{Casalderrey}, while the
parton cascade model works near and far way from jet region
harmoniously. Since only two-body scatterings are included in the
current AMPT model, we conclude that our results are due to the
big partonic interaction cross sections which can transit the
energy of high-$p_{T}$ parton into a hydro-like pattern that stems
from successive parton interactions.

We have not considered the effect from hadronic rescattering on
Mach-like structure in the present work. However, it has been
found hadronic rescattering is also important for Mach-like
structure in our previous studies\cite{di-hadron,three-particle}.
We found that the effect of hadronic rescattering is weak for
di-hadron correlations but considerable for three-particle
correlations, for this $p_{T}$ window cut. Fortunately, hadronic
rescattering can not erase the correlations from a partonic shock
wave, therefore the observed Mach-like structure in final state
can carry the information about partonic Mach-like shock wave.

In conclusion, we used a partonic transport model to study the
production mechanism of Mach-like structure with two- and
three-particle correlation methods. Our results indicate that
parton cascade mechanism can couple many partons together to
exhibit a partonic Mach-like shock wave by strong partonic
interactions. The partonic Mach-like shock wave is born at $\sim$
1.5 fm/c and takes stable shape around $\sim$ 5 fm/c, and the
birth of Mach-like shock wave can boost its own growth quickly by
further parton cascade. Since the formation and development of
Mach-like shock wave are sensitive to the strength of partonic
interactions and the lifetime of partonic matter, the studies of
Mach-like structure could be helpful to explore the characters of
partonic interactions and the critical properties of reaction
system, such as the lifetime of partonic matter and hadronization
time.

This work was supported in part by the National Natural Science
Foundation of China  under Grant No 10535010 and 10135030, and
the Shanghai Development Foundation for Science and Technology
under Grant Numbers 05XD14021 and 06JC14082.

\end{document}